\documentclass[12pt]{article}
\usepackage{graphicx,amsmath}
\usepackage{units}

\newlength{\dinwidth}
\newlength{\dinmargin}
\def\lapproxeq{\lower .7ex\hbox{$\;\stackrel{\textstyle                                                    
<}{\sim}\;$}}                                                    
\def\gapproxeq{\lower .7ex\hbox{$\;\stackrel{\textstyle                                                    
>}{\sim}\;$}}                                                    
\def\be{\begin{equation}}                                                    
\def\ee{\end{equation}}                                                    
\def\bea{\begin{eqnarray}}                                                    
\def\eea{\end{eqnarray}}

\def\qq{q\bar{q}}

\def\sh{\hat s}
\def\sh2{{\hat s}^2}
\def\PDF{{\rm PDF}}
\def\CLO{C^{\rm LO}}
\def\CNLO{C^{\rm NLO}}

\def\s{\hat{s}}
\begin{document}

\begin{flushright}                                                    
IPPP/12/38  \\
DCPT/12/76 \\                                                    
\today \\                                                    
\end{flushright} 

\vspace*{0.5cm}

\begin{center}
{\Large \bf The treatment of the infrared region \\
 \vspace{0.5cm}
in perturbative QCD}

\vspace*{1cm}
                                                   
E.G. de Oliveira$^{a,b}$, A.D. Martin$^a$ and M.G. Ryskin$^{a,c}$  \\                                                    
                                                   
\vspace*{0.5cm}                                                    
$^a$ Institute for Particle Physics Phenomenology, University of Durham, Durham, DH1 3LE \\                                                   
$^b$ Instituto de F\'{\i}sica, Universidade de S\~{a}o Paulo, C.P.
66318,05315-970 S\~{a}o Paulo, Brazil \\
$^c$ Petersburg Nuclear Physics Institute, NRC Kurchatov Institute, Gatchina, St.~Petersburg, 188300, Russia \\          
                                                    
\vspace*{1cm}                                                    
                                                    
\begin{abstract}                                                    

We discuss the contribution coming from the infrared region to NLO matrix elements and/or coefficient functions of hard QCD processes. Strictly speaking, this contribution is not known theoretically, since it is beyond perturbative QCD. For DGLAP evolution all the infrared contributions are collected in the phenomenological input parton distribution functions (PDFs), at some relatively low scale $Q_0$; functions which are obtained from a fit to the `global' data. However dimensional regularization sometimes produces a non-zero result coming from the infrared region. Instead of this conventional regularization treatment, we argue that the proper procedure is to first subtract from the NLO matrix element the contribution already generated at the same order in $\alpha_s$ by the LO DGLAP splitting function convoluted with the LO matrix element. This prescription eliminates the logarithmic infrared divergence, giving a well-defined result which is consistent with the original idea that everything below $Q_0$ is collected in the PDF input. We quantify the difference between the proposed treatment and the conventional approach using low-mass Drell-Yan production and deep inelastic electron-proton scattering as examples; and discuss the potential impact on the `global' PDF analyses. We present arguments to show that the difference cannot be regarded as simply the use of an alternative factorization scheme.

\end{abstract}                                                        
\vspace*{0.5cm}                                                    
                                                    
\end{center}

\section{Introduction}

While studying the perturbative QCD (pQCD) description\footnote{A preliminary study can be found in \cite{OMR}.} of the Drell-Yan production of low-mass $\mu^+\mu^-$ pairs, in preparation for the interpretation of the forthcoming LHC measurements\footnote{See, for example, the preliminary LHCb data \cite{LHCbDY}.}, we unearthed a puzzle in the conventional procedure to remove the infrared divergencies. The puzzle is not just confined to the description of Drell-Yan production, but occurs for other QCD processes, such as deep inelastic electron-proton scattering, etc. The resolution of the puzzle can lead to numerical corrections at low scales. As such, it has the potential to influence the global PDF analyses \cite{MSTW}-\cite{JR}.

To introduce the problem, recall that pQCD calculations are based on factorization theorems. The cross section of a hard process is calculated from the convolutions of parton distribution functions (PDFs) with the cross sections of the various hard subprocesses.  The PDFs satisfy DGLAP evolution in ln$Q^2$ starting from an initial input at some low starting scale $Q^2_0$. The input values are known from `global' fits to all the available deep inelastic and related hard scattering data. Below $Q_0$ the higher-order $\alpha_s$ and higher-twist and other corrections, become too large for the use of pQCD.

In the collinear approximation, at leading order (LO), each additional power of small $\alpha_s$ is compensated by the large logarithm of a hard scale (such as ln$Q^2$). At next-to-leading order (NLO) we have an $\alpha_s$ term in the hard matrix element and/or the DGLAP splitting function unaccompanied by a large log. At NNLO the $\alpha_s^2$ terms do not have large logs, and so on.

There is no problem with this procedure at LO --- due to strong $k_t$-ordering, when the DGLAP evolution is started from $Q_0$, we do not encounter contributions from the $k_t<Q_0$ domain. On the other hand, already inside the NLO cell there appears a contribution with $k_t<Q_0$. How to treat the (infrared) contribution from this large distance domain is the subject of this note.

We discuss two alternative treatments of the infrared domain. First, we adopt a practical approach, which we shall call the `physical' prescription, see Section \ref{sec:PA}. To avoid double counting, we have to subtract from the NLO diagram the $\alpha_s*$log contribution that has already been generated by LO DGLAP evolution. As we will show, this completely removes the infrared divergency from the matrix element.  To be precise, after the subtraction, there remains a very small contribution coming from the $k_t<Q_0$ domain of ${\cal O}(Q_0^2/\mu_F^2)$, where $\mu_F$ is the factorization scale, which we will discuss in Section \ref{sec:PA}.  

The second treatment of the infrared divergency, which we will call
 the `conventional' prescription, is based on dimensional regularisation, see Section \ref{sec:CA}. Here just the $1/\epsilon$ pole, which corresponds to the anomalous dimension generated by LO DGLAP evolution, is subtracted from the NLO contribution. In this case, after the subtraction, we are left with a contribution from the $k_t<Q_0$ region
 which does not vanish as $Q_0^2 \to 0$.  So, surprisingly, the two treatments of the infrared domain yield different results.
 
Sections \ref{sec:PA} and \ref{sec:CA} introduce the two treatments in more detail, using the NLO coefficient function of Drell-Yan production as an example. Then in Section 4 we quantify the difference of the contributions from the infrared domain taking again the NLO coefficient function in Drell-Yan production and, in addition, the $\gamma^*g$ coefficient function, $C_g$, of deep inelastic scattering, as topical examples.  Section \ref{sec:discuss} (as well as the latter part of Section 4) discusses the origin of the difference.

\section{Calculation at NLO -- the physical approach   \label{sec:PA}}

Schematically, we may write the cross section in the form
\be
\label{eq:sig}
d\sigma/d^3p~=~\int dx_1dx_2~\PDF(x_1,\mu_F)~|{\cal M}(p;\mu_F,\mu_R)|^2~\PDF(x_2,\mu_F)\ ,
\ee
where a sum over the various pairs of PDFs is implied. 
The matrix elements squared, $|{\cal M}|^2$, describe the cross sections of the elementary partonic subprocesses. 
However, the problem is that using 
(\ref{eq:sig}) 
we do not know the factorization scale $\mu_F$ at which the PDFs are measured. Moreover, in the low $x$ region, the PDFs strongly depends on the choice of $\mu_F$.

In general, after the summation of all perturbative orders, the final result should not depend on the choice of $\mu_F$ that is used to separate the incoming PDFs from the hard matrix element. Thus we need to account for the NLO, NNLO,...corrections.
Contributions with 
low virtuality, $q^2<\mu_F^2$, of the incoming partons are included in the PDFs, while those with 
$q^2>\mu_F^2$ are assigned to the matrix element.
 
Let us start with the LO expression for the cross section. In the collinear approach, the cross section has the form
\be
\sigma(\mu_F)~=~\PDF(\mu_F)\otimes\CLO \otimes \PDF(\mu_F). 
\ee
The effect of varying the scale from $m$ to $\mu_F$, in both the left and right PDFs, can be expressed, to first order in $\alpha_s$, as
\be 
\sigma(\mu_F)=\PDF(m)\otimes    
 \left(\CLO ~+~\frac{\alpha_s}{2\pi}{\rm ln}\left(\frac{\mu_F^2}{m^2}\right)(P_{\rm left}\CLO+\CLO P_{\rm right})\right)\otimes \PDF(m),
\label{eq:5}
\ee
where the splitting functions $P_{\rm right}$ and $P_{\rm left}$
act on the right and left PDFs respectively. 
Recall that in calculating the $\alpha_s$ correction in (\ref{eq:5}),
the integral over the transverse momentum (virtuality) of the parton in the LO DGLAP evolution was approximated by the pure logarithmic $dk^2/k^2$ form. That is, in the collinear approach, the Leading Log Approximation (LLA) is used.

\begin{figure} [htb]
\begin{center}
\vspace{-1cm}
\includegraphics[height=8cm]{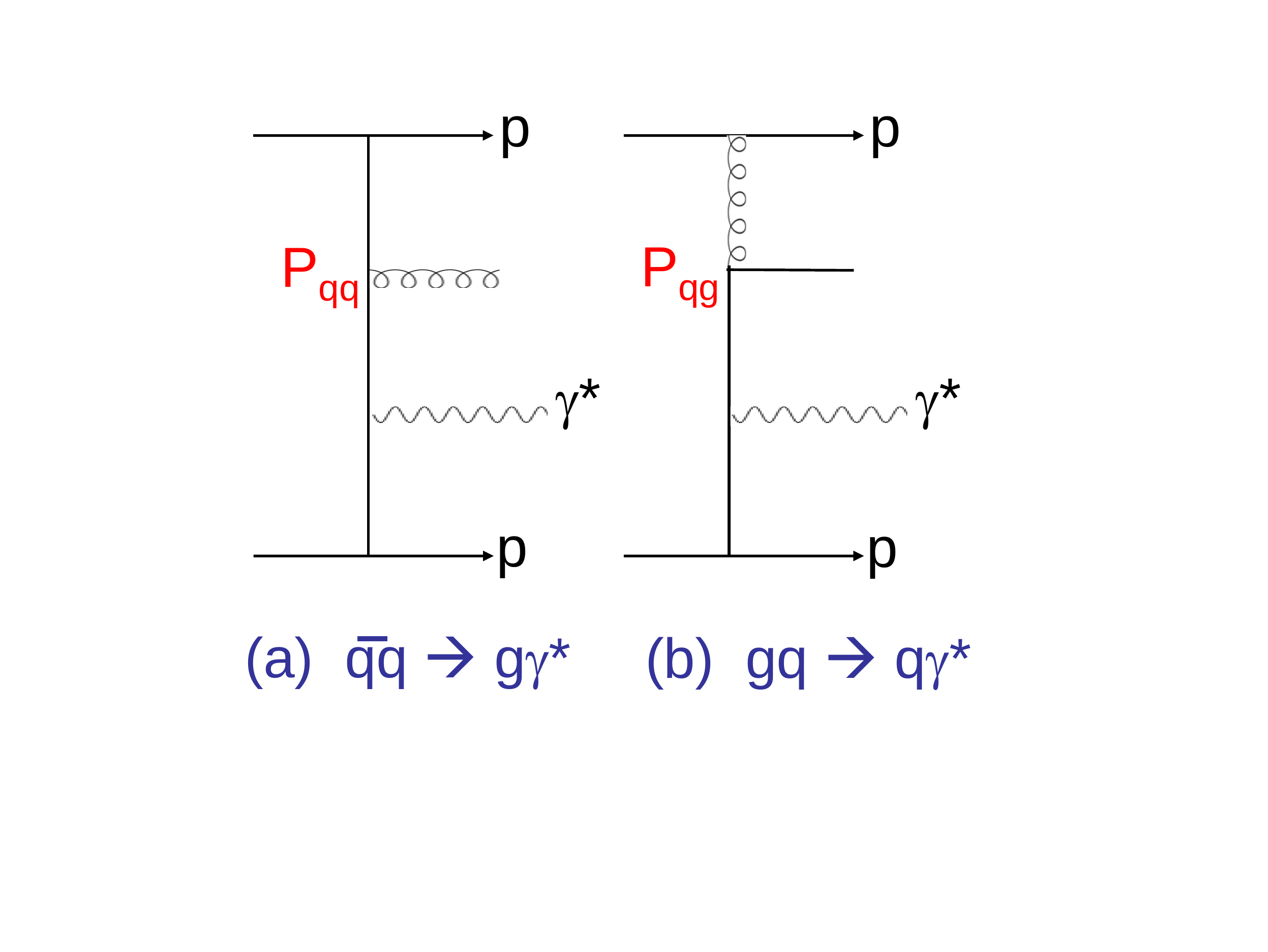}
\vspace{-2cm}
\caption{\sf Diagrams (a,b) show NLO subprocesses for Drell-Yan production resulting from the splitting of the upper PDF (or `right' PDF, in the notation of (\ref{eq:5})). }
\label{fig:fd}
\end{center}
\end{figure}
Now we turn to the expression for the cross section at NLO. First, we note that the original Feynman integrals corresponding to the NLO matrix element, $\CNLO$, do not depend on $\mu_F$. However, we will see below how a scale dependence enters.  At NLO we may write
\be
\sigma(\mu_F)~=~\PDF(\mu_F)\otimes(\CLO + \alpha_s \CNLO_{\rm corr}) \otimes \PDF(\mu_F), 
\label{eq:stab}
\ee
where we include the NLO correction to the coefficient function.  However to avoid double counting we must subtract the DGLAP-generated $\alpha_s$ term of (\ref{eq:5}). This cancels the infrared singularity, and gives a non-singular result which can be safely integrated in normal 4-dimensional space.

\subsection{Example: Drell-Yan production at NLO} 

We illustrate the general procedure in terms of the specific example of Drell-Yan production of a $\mu^+\mu^-$ pair of mass $M$. At LO the subprocess is $q\bar{q} \to \gamma^*$, while at NLO we have the $2\to 2$ subprocesses $\qq \to g\gamma^*$ and  $gq \to q\gamma^* $ shown in diagrams (a,b) of Fig. \ref{fig:fd}. These NLO contributions are now calculated with better, than LLA, accuracy. However part of these contributions are already included, to LLA accuracy, in the second term in (\ref{eq:5}),  where the splitting functions
\begin{equation}
P_{\rm right}=P_{qq}+P_{qg}~,~~~~~~~~~P_{\rm left}=P_{\bar{q}\bar{q}}+P_{\bar{q}g}
\end{equation}
act on the right and left PDFs respectively\footnote{ We may equally well have incoming $\bar{q}$'s in $P_{\rm right}$ and incoming $q$'s in $P_{\rm left}$.}.
So this LLA part should be subtracted from $\CNLO$. 
 
 For Drell-Yan production at low $x$, the majority of quarks/antiquarks are produced via the low-$x$ gluon-to-quark splitting, $P_{qg}$. That is, the most important NLO subprocess is $gq\to q\gamma^* $.
The corresponding cross section, for the production of a $\mu^+\mu^-$ pair of mass $M$ and rapidity $Y$, has the schematic form
\begin{equation}
\frac{d\sigma}{dM^2dYdt}=\int dx_1dx_2~{\rm PDF}_i(x_1)~\frac{d\hat{\sigma}_{ij}}{dt}~{\rm PDF}_j(x_2)~~\delta(x_1^\gamma x_2^\gamma s-M^2)~~\delta \left(\frac{1}{2} {\rm ln}\frac{x^\gamma_1}{x^\gamma_2}-Y\right),
\end{equation}
where $x^\gamma_i=x_i$ in the LO case, but $x^\gamma_i$ are the true momentum fractions ($x^+$ and $x^-$) carried by the photon at NLO. The subprocess cross sections have the form
\begin{equation}
\hat{\sigma}_{ij}~=~\hat{\sigma}^{\rm LO}_{ij}~+~\hat{\sigma}^{\rm NLO}_{ij}~+~...,
\end{equation}
where, for the main NLO subprocess, we have
\begin{equation}
\frac{d \hat{\sigma}_{qg}}{dt} = \frac{1}{9} \frac{\alpha^2 \alpha_s}{\s^2} \left[-\frac{\s}{t} - \frac{t}{\s} - \frac{2M^2 u}{\s t} \right].
\label{eq:nlo}
\end{equation}
Since $M^2 = z \s$ and $u=M^2-\hat s -t$, this gives  
\begin{equation}
\frac{d \hat{\sigma}(gq\to q\gamma^* )}{d |t|}=\frac{\alpha^2 \alpha_s z}{9 M^2} \frac{1}{|t|} \left[( (1 - z)^2 + z^2) + z^2 \frac{t^2}{M^4} - 2 z^2 \frac{t}{M^2} \right].
\label{eq:nlo2}
\end{equation}

In order to calculate the inclusive cross section $d {\sigma}/d M^2$, it seems that we have to integrate over $t$ starting from $t=0$. If this were necessary, then we would face an infrared divergency.

\subsection{Treatment of the infrared region in the physical approach}

When dealing with PDFs we avoid the problems of confinement and interactions at large distances. We start with some phenomenological input at a relatively large $Q_0$, that is $Q_0^2\gg\Lambda^2_{\rm QCD}$, and consider just the evolution of the PDFs with $Q^2$ increasing from $Q_0^2$ to the factorization scale $\mu_F^2$. Everything below $Q_0$ is absorbed in the phenomenological input PDF. Accepting this logic, it is natural to replace the lower limit $t=0$ by the same $Q_0^2$, and never to consider the contribution from low virtualities, $k^2<Q_0^2$.

The subtraction of the LO DGLAP contribution from the NLO $d\hat{\sigma}/dt$  completely eliminates the $1/t$ singularity as $t \to 0$ by introducing a $\Theta(|t|-\mu^2_F)$ function in the first term of (\ref{eq:nlo2}). Let us explain, more explicitly, how this happens. We denote the  $gq \to q\gamma^*$  ${\cal O}(\alpha_s)$ contribution by $\hat{\sigma}^{(1)}$.  The result of the explicit calculation of the $gq \to q\gamma^*$ cross section is shown in (\ref{eq:nlo2}). On the other hand, this contribution may be written as the term generated by LO DGLAP evolution and the remaining NLO part, that is
\begin{equation}
d\hat\sigma^{(1)}~=~d\hat\sigma^{\rm NLO}_{\rm rem}~+~\sigma^{\rm LO}_{\bar qq}\otimes
\frac{\alpha_s}{2\pi}P^{\rm LO}_{\bar qg}dt.
\label{eq:nlo7}
\end{equation}
The last term reads
\begin{equation}
\sigma^{\rm LO}_{\bar qq}\otimes\frac{\alpha_s}{2\pi}P^{\rm LO}_{\bar qg}dt ~=~
\frac{\alpha^2\alpha_s z}{9M^2}\frac{dt}{t}[ (1 - z)^2 + z^2] ~\Theta(\mu_F^2-|t|),
\label{eq:nlo5}
\end{equation}
where here DGLAP evolution has accounted for all virtualities $|t|=k^2<\mu_F^2$; and where the contribution of $k^2<Q^2_0$ is hidden in the phenomenological input PDF. After the subtraction of this LO DGLAP generated term, the remaining contribution of (\ref{eq:nlo2}) is 
\begin{equation}
\frac{d \hat{\sigma}^{\rm NLO}_{\rm rem}}{d |t|}=\frac{\alpha^2 \alpha_s z}{9 M^2} \frac{1}{|t|} \left[[ (1 - z)^2 + z^2] ~\Theta(|t|-\mu_F^2) ~+~ z^2 \frac{t^2}{M^4} ~-~ 2 z^2 \frac{t}{M^2} \right].
\label{eq:nlo3}
\end{equation}
which has no singularity as $t \to 0$. 

Recall that we {\it need} to subtract the DGLAP generated contribution of (\ref{eq:nlo5}) in order to avoid the double counting. This subtraction is unambiguously defined. It is done in terms of $t$ (and not $k_t$) since the original DGLAP evolution was written as an evolution in $k^2=-t$, and not as an evolution in $k_t$.

\begin{figure} [htb]
\begin{center}
\includegraphics[height=8cm]{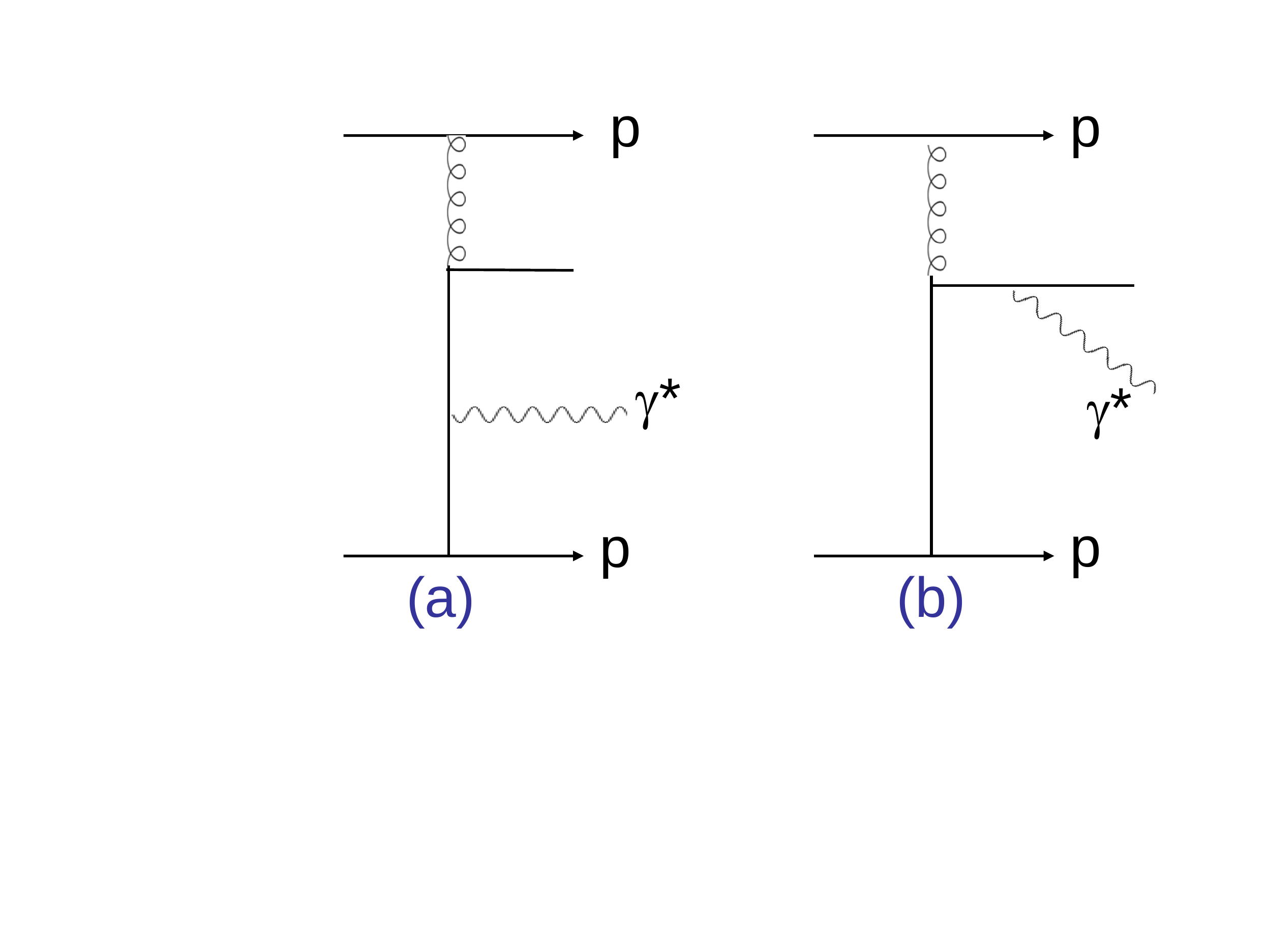}
\vspace{-3cm}
\caption{\sf (a) Ladder and (b) non-ladder diagrams. }
\label{fig:lnl}
\end{center}
\end{figure}

A small uncertainty is still present coming from the treatment of the non-singular terms in (\ref{eq:nlo3}). We return to the choice of the lower limit of integration.  It is not evident whether we should integrate from $t=0$ or from $|t|=Q_0^2$. From the ladder diagram, Fig.~\ref{fig:lnl}(a), which is of exactly the same form as that in LO DGLAP evolution in which everything below $Q_0^2$ is absorbed in the input, we are led to integrate from $|t|=Q_0^2$. However, the NLO hard subprocess cross section also includes non-ladder diagrams, Fig.~\ref{fig:lnl}(b) with an $s$-channel quark, for which the $|t|<Q_0^2$ domain does not correspond to low virtuality. Thus the true uncertainty due to the non-singular terms is ${\cal O}(Q_0^2/\mu_F^2)$, which may be neglected for $\mu_F \gg Q_0^2$. 

\section{Calculation at NLO -- conventional approach  \label{sec:CA}}

We now consider the conventional prescription for the evaluation of
\begin{equation}
\frac{d \hat{\sigma}(gq\to q\gamma^* )}{d |t|}=\frac{\alpha^2 \alpha_s z}{9 M^2} \frac{1}{|t|} \left[( (1 - z)^2 + z^2) + z^2 \frac{t^2}{M^4} - 2 z^2 \frac{t}{M^2} \right].
\label{eq:nloconv}
\end{equation}
Recall, that to calculate the inclusive cross section $d {\sigma}/d M^2$, we appear to have to integrate over $t$ starting from $t=0$, and that we face an infrared divergency\footnote{In perturbation theory this divergency is avoided by the (small) current quark mass, $m_q$. However, in practice, the light quark mass is neglected.}.  In the conventional approach the integral is regularized by introducing the $4+2\epsilon$ dimensional space \cite{Alt,Alt2,ESW}. Then the contribution from very small $t$ produces a $1/\epsilon$ pole, which is absorbed into the incoming PDF. However, in $4+2\epsilon$ dimensional space, the number of gluon states with transverse polarisation is $2+2\epsilon$, giving $\epsilon$ in the numerator of the matrix element. Besides this, there is an $\epsilon$ dependence coming from the decomposition of the phase space factor, like $(1-z)^{-\epsilon}$.
So finally we have an $\epsilon/\epsilon$ term, which produces a non-zero result as $\epsilon \to 0$.

Simultaneously, we have to consider the same diagram generated by LO DGLAP evolution, which gives a $1/\epsilon$ pole. 
 The $1/\epsilon$ poles cancel, but we are left with the $\epsilon/\epsilon$ term, which produces a non-zero result as $\epsilon \to 0$.   Unlike the physical approach, this non-zero term does not vanish as ${\cal O}(Q_0^2/\mu_F^2)$, and does not vanish for $Q_0^2 \ll \mu_F^2$.

Thus, suprisingly, we find that the physical and conventional  prescriptions yield different results. We quantify this difference in the next Section, and then in Section \ref{sec:discuss} we discuss its origin.

Sometimes the exact treatment of the infrared singularity does not matter. In particular, the emission of a soft gluon has a logarithmic infrared divergency which, according to the Bloch-Nordsieck theorem, is exactly cancelled between the real emission contribution and the virtual loop correction\footnote{This is the justification for the $+$ prescription used for the $1/(1-z)$ terms in the DGLAP splitting functions.}. However, in NLO Drell-Yan or in the $\gamma g$ coefficient function for the structure function $F_2$ of deep inelastic scattering, for example, the difference in the treatment of the infrared singularity does matter.

\section{Quantitative estimates  \label{sec:ex}}

To illustrate the numerical differences of the two treatments of the infrared contribution, we study, as examples, Drell-Yan production and the structure function $F_2$ of deep inelastic scattering.

\subsection{Drell-Yan}

\begin{figure} [htb]
\begin{center}
\includegraphics[height=8cm]{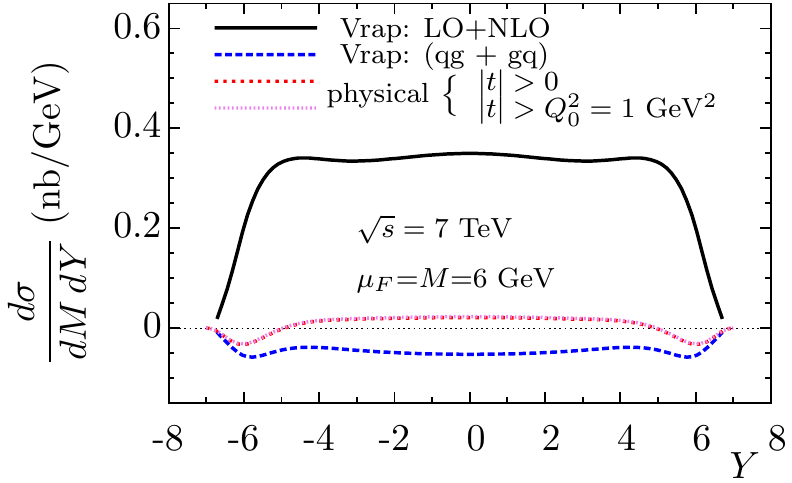}
\caption{\sf The LO+NLO cross section for Drell-Yan production of a $\mu^+\mu^-$ pair of mass $M=6$ GeV and rapidity $Y$ at pp collider energy of 7 TeV, obtained using MSTW NLO PDFs \cite{MSTW}. The two dashed curves correspond to the two different treatments of the infrared singularity in the NLO contribution.  The lower dashed curve is obtained using `conventional' $\epsilon$ regularisation, whereas the upper dashed curve is obtained using the `physical', approach working in normal 4 dimensional space.  For the later we show the result for $|t|>0$ (short-dashed) and imposing a cut $|t|>1~{\rm GeV}^2$ (dotted) --- the minute difference is invisible on the plot. }
\label{fig:DY}
\end{center}
\end{figure}

The Drell-Yan cross section calculated using the Vrap code \cite{Vrap} and MSTW08 NLO PDFs \cite{MSTW} is shown by the continuous curve in Fig.~\ref{fig:DY} for the LHC energy $\sqrt{s}=7$ TeV and a mass $M=6$ GeV of the produced $\mu^+\mu^-$ pair. As usual, the factorization scale is chosen to be $\mu_F=M$. The long-dashed line is the remaining contribution of the NLO $gq \to q\gamma^*$ subprocess calculated in the framework of dimensional regularisation. We see that it turns out to be negative, since, for $\mu_F=M$, the contribution formally generated by LO evolution, (\ref{eq:nlo5}), is larger than the whole ${\cal O}(\alpha_s)$ cross section.

On the other hand, if we were to adopt the `physical' approach and subtract the explicit DGLAP-generated term (as discussed in Section 3), then we obtain the result shown by the short-dashed curve, which is significantly different to that obtained using  dimensional regularisation.
Moreover, we see the uncertainty coming from the $|t|<Q_0^2$ domain is practically invisible in this figure --- essentially identical results are obtained by integrating the last two (non-singular) terms in [...] in (\ref{eq:nlo3})
with a lower limit $|t|=Q_0^2=1~{\rm GeV}^2$ (dotted curve) and with $t=0$ (short-dashed curve).

\subsection{Deep inelastic scattering}

\begin{figure} [htb]
\begin{center}
\includegraphics[height=8cm]{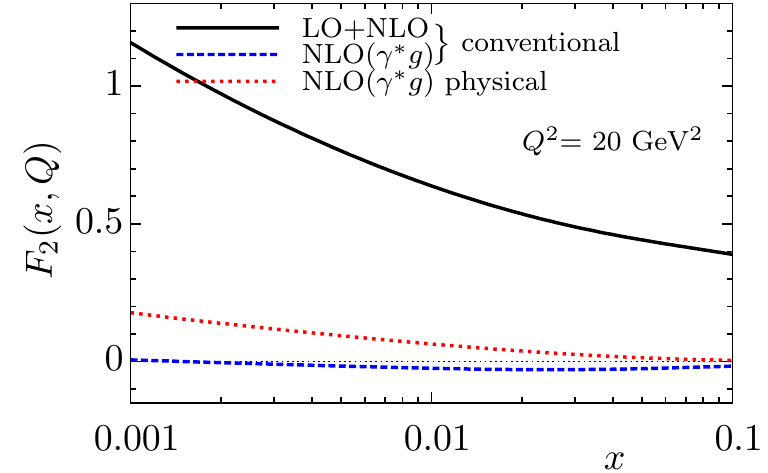}
\caption{\sf The continuous curve is the LO+NLO result for the structure function $F_2(x,Q^2)$ for electron-proton DIS obtained from MSTW08 NLO PDFs \cite{MSTW}. The long-dashed and short-dashed curves are the NLO contributions arising from photon-gluon fusion ($\gamma^*g \to q\bar{q}$) for the `conventional' and the `physical' treatment of the infrared $1/t$ divergence, respectively. }
\label{fig:DIS}
\end{center}
\end{figure}

\begin{figure} [htb]
\begin{center}
\includegraphics[height=8cm]{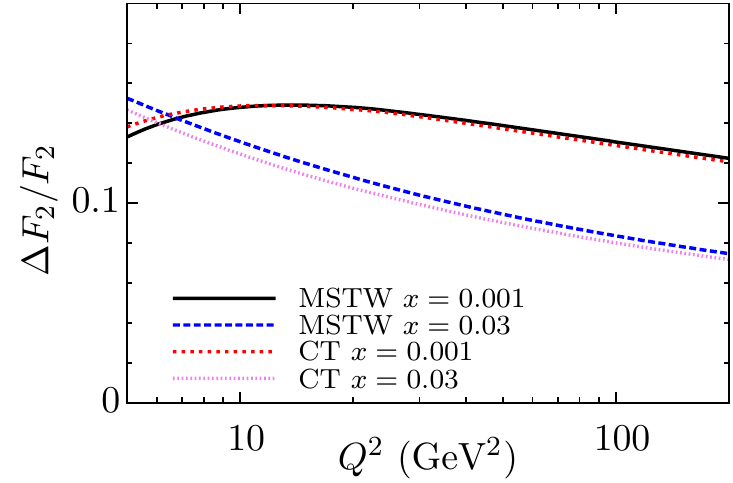}
\caption{\sf The fractional correction to $F_2$ arising from using the `physical', rather the `conventional', treatment of the infrared divergency in the $\gamma^*g$ coefficient function, shown as  function of $Q^2$ for two values of $x$, and for two different sets of NLO PDFs, namely  MSTW08 \cite{MSTW} and CT10 \cite{CT10}. }
\label{fig:DISQ}
\end{center}
\end{figure}

    An analogous situation is shown in Fig.~\ref{fig:DIS} and  \ref{fig:DISQ} for the NLO gluon contribution to $F_2$ in DIS. The calculation is
very similar to that for the Drell-Yan case since these two subprocesses are
closely related by crossing symmetry. As before, the explicit subtraction of the term generated by the LO DGLAP evolution leads to a result for the $\gamma^*g$ coefficient function,
\begin{equation}
C_g~=~T_R\left([(1-z)^2+z^2]~{\rm ln}\frac{1}{z}~+~6z(1-z)~-~1\right),
\label{eq:13}
\end{equation}    
     that is different to that obtained by the conventional prescription,
\begin{equation}
C_g~=~T_R\left([(1-z)^2+z^2]~{\rm ln}\frac{1-z}{z}~+~8z(1-z)~-~1\right).
\label{eq:14}
\end{equation} 
Here the notation of \cite{ESW} has been used.  The additional ln$(1-z)$ singularity reflects the threshold singularity; as $z \to 1$ the phase space available for the quark and antiquark is proportional to $(1-z)$.  In $\epsilon$ regularization this threshold singularity produces the factor 
\begin{equation}
(1-z)^{-\epsilon}=1-\epsilon\ln(1-z)+O(\epsilon^2),
\end{equation}
which, on being multiplied by the $1/\epsilon$ pole, gives a constant   $\ln(1-z)$ term, that does not vanish as $\epsilon\to 0$.
Unlike the conventional treatment, the physical prescription has, at most, only a very small ${\cal O}(Q^2_0/\mu_F^2)$ infrared contribution coming from the non-singular terms in the infrared region.

To gain insight into the difference between the `conventional' and `physical' treatments of the infrared divergency, we discuss the $t$-dependence of the $\gamma^* g \to q\bar{q}$ matrix element. Adopting the `physical' procedure we start with the explicit perturbative QCD result and subtract the LO DGLAP-generated contribution. The result, after the photon polarizations have been convoluted with $g_{\mu\nu}$, may be written in the form
\begin{equation}
|{\cal M}|^2~=~|{\cal M}_t|^2~+~|{\cal M}_u|^2,
\label{eq:15}
\end{equation}
where\footnote{Integrating (\ref{eq:16}) over $t$, and adding the term which originates from $F_L$ (where
there is no infrared divergency), leads to (\ref{eq:13}).}
\begin{equation}
|{\cal M}_t|^2~=~[(1-z)^2+z^2] \frac{Q^2}{z t} \Theta(|t|-\mu_F^2) ~-~1.
\label{eq:16}
\end{equation}
Similarly for ${\cal M}_u$ with $u$ interchanged with $t$.  The result is non-singular as $t \to 0$.  Numerically the non-singular contribution coming from the $0\le |t|\le Q_0^2$ domain is not important\footnote{For the analogous Drell-Yan process we  saw in Fig.~\ref{fig:DY} that the difference between the results obtained by integrating over the regions $|t|>0$ and $|t|>Q_0^2=1~{\rm GeV}^2$ were invisible on the plot.  The same insensitivity of the results, when using the `physical' treatment of infrared domain,  applies also to the short-dashed curve in the $F_2$ plot of Fig~\ref{fig:DIS}.}.
After integration over $t$, this contribution vanishes as ${\cal O}(Q^2_0/\mu^2_F)$ as $Q_0 \to 0$.   Indeed, the `physical' prescription eliminates the logarithmic infrared divergence, giving a well-defined result which is consistent with the original idea that everything below $Q_0$ is collected in the PDF input. 

On the other hand, the $1/t$ singularity in $|{\cal M}_t|^2$, in the `conventional' treatment, is first regularized by working in $4+2\epsilon$ space. After the subtraction of the analogous DGLAP-initiated contribution, the $1/\epsilon$ term is cancelled. However, the resulting $\infty-\infty$ subtraction leaves some non-zero remaining contribution, which can be viewed as effective non-local vertices, which act as a $\delta (t)$ contribution. Recall that  analogous vertices arising from an ultra-violet divergence, play the role of {\it local} counter terms. Dimensional regularization provides us with the possibility to introduce such counter terms in a way which does not spoil the renormalizability of the theory. However, the analogous effective vertices of infrared-origin are unacceptable. They are contrary to the confinement property of QCD. Due to the confinement property it is not possible to continue the $1/t$ behaviour below some low scale of ${\cal O}(\Lambda_{\rm QCD})$. Recall that in the global PDF analyses this low $t$ contribution is included phenomenologically in the input PDFs at the starting scale $Q_0$, which is taken to be larger than $\Lambda_{\rm QCD}$.

\section{Discussion  \label{sec:discuss}}

Our approach has some similarities with the `subtraction method' \cite{catani,submethod} where to regularize the infrared divergency one
subtracts from the real cross section of $n+1$ parton production, $d\sigma^R_{n+1}$, a term $d\sigma^A$,  which has exactly the same divergency.  Simultaneously the $d\sigma^A$ term is added to the virtual loop correction for the $n$ parton cross section, $d\sigma^V_n$. After this both the real and the virtual contribution have no infrared divergency, and so may be
calculated in the normal 4 dimensional space (without the $\epsilon$ decomposition).

 In comparison with this `subtraction method', where the subtraction term is introduced artificially, here we subtract the contribution {\it really} generated by  DGLAP evolution. On one hand, this guarantees the absence of double counting, while
 on the other hand, it eliminates the infrared divergency.  Note, also, that in our particular examples, like the NLO correction to $F_2$ due to  photon-gluon fusion, or the NLO correction to the Drell-Yan cross section due to $gq\to q\gamma^*$ subprocess, there is no analogous LO contribution caused by gluons.  Therefore there is no NLO virtual correction, like $d\sigma^V_n$,  in the `subtraction method'.

Let us return to the comparison of the two treatments of the infrared domain discussed in this paper: that is the `conventional' $\epsilon$ regularisation subtraction and the `physical' direct DGLAP subtraction methods. Why do the results differ? 
The reason is subtle.  In the `physical' approach we do not need to assume that perturbative QCD is valid for virtualities $k^2<Q^2_0$, and, in particular, in the confinement domain.
 We evolve starting from some phenomenological `input' distribution at $\mu^2=Q^2_0$. By considering different choices of $Q_0^2$ we verify that the contribution coming from the  $k^2<Q^2_0$ domain is small, and 
 vanishes as $Q_0^2 \to 0$. On the other hand, the conventional dimensional $\epsilon$ regularisation procedure continues the $1/t$ singular behaviour below $\Lambda_{\rm QCD}^2$, giving finally a non-vanishing infrared contribution, which may be viewed as a $\delta(t)$ term during the $t$ integration.  However, the part coming from low virtualities, $k^2<Q^2_0$, has no reason to have perturbative form, and, moreover, does not vanish as $Q_0^2 \to 0$. 

We elaborate this in a little more detail.  Dimensional regularization is clearly a valuable tool in the ultraviolet region. But, here, we are concerned about its use in the infrared domain.
The hope was that for infrared stable quantities (like $F_2$) the result will not depend on the contribution from the large distance domain; the real and virtual loop contributions will cancel each other. Sometimes this is true. In particular, due to the
Bloch-Nordsieck theorem, we can use the so-called `plus prescription'
which eliminates the $1/(1-z)$ singularity in splitting functions.

However, there are examples where `conventional'
 dimensional regularization gives a non-zero infrared contribution 
 of the form of $\epsilon/\epsilon$, which is the correct mathematical result of
a particular perturbative QCD diagram calculation, but which makes no physical sense for real QCD with confinement.  Here 
we have presented two examples (the NLO $\gamma^*g$ coefficient function in Drell-Yan production and $C_g$ in DIS) and have shown that
the subtraction of the LO DGLAP-generated contribution completely eliminates the infrared singularity; the cancellation occurs between terms in $\CNLO$ and in the convolution
the $\CLO \otimes P^{\rm LO}$. Recall that this subtraction is necessary to avoid double counting. After this, the integral can be calculated in normal 4-dimensional space, and, surprisingly,
as demonstrated above, the results differ from those found in the `conventional' approach.

Similarly, using the conventional $\epsilon$ regularisation procedure, the NLO splitting functions, in particular $P_{gg}^{\rm NLO}(z)$, of DGLAP evolution may also contain  contributions of infrared origin which are inconsistent with confinement.
On the other hand, in the `physical' approach, the absence of the infrared 
singularity is due to the cancellation of the contribution to 
the NLO splitting function arising from the 
direct calculation of the NLO Feynman diagram and the contribution {\it already} 
generated by LO DGLAP evolution. That is, it is due to cancellations of the contributions occurring in 
$P^{\rm NLO}$ and in $P^{\rm LO}\otimes P^{\rm LO}$.   A more detailed discussion is given in the Appendix.
The derivation of these new infrared-corrected NLO splitting functions is more complicated than the treatment of the coefficient functions that we have used as examples, but it is still possible\footnote{We are at present tackling this problem.}. 

In summary, the problem cases for the `conventional' treatment of the infrared divergence are the NLO coefficient and splitting functions, in particular the $C_g$ and $P_{gg}$ functions, which leave `erroneous' non-singular contributions. The main effect will be on the behaviour of the gluon at low scales, a domain where the present global parton analyses show unexpected `valence-like' $x$ behaviour of the gluon; see, for example, \cite{MSTW,CT10,NNPDF21}.
In order to carry about such a `global' investigation it will be necessary to compute, not only the coefficient functions, but also the NLO splitting functions using the `physical' treatment of the infrared domain.

One might think, say, in the DIS example that we considered, that the infrared corrections to the coefficient and splitting functions could compensate each other, and that the transition from the `conventional' coefficient function (which includes a non-vanishing infrared contribution inconsistent with colour confinement) to that obtained in a `consistent' treatment, may be effectively accounted for by a re-definition of the PDFs, and thus considered as an alternative `factorization scheme'. 
However such an alternative scheme, which may correspond to new coefficient functions, prescribes a new set of splitting functions to describe the evolution of the re-defined PDFs, which are, in general, different to the infrared-corrected splitting functions.  

Even without explicit knowledge of the NLO splitting functions corresponding to the `physical' treatment of the infrared domain,
there are several ways to see that compensation is not possible and that the results {\it cannot} be regarded as simply a scheme change.  A hint, that this is so, is the presence of the additional ln$(1-z)$ singularity in the `conventional' $C_g$ of (\ref{eq:14}).
An argument is to consider the non-singlet channel. Here, due to flavour conservation, the PDF normalization is fixed, and we have no possibility to re-define the non-singlet PDF via the admixture with some gluon or other quark flavour contribution. In other words, there is no chance to compensate the correction to NLO coefficient function by the re-definition of the (non-singlet) parton density and/or by the correction
to the NLO splitting function, which is exponentiated, that is iterated many times already at NLO, especially at small $x$. It is not a NNLO effect. 

Alternatively, to be more explicit, we may consider the quark contribution to the $gg$ splitting function. This is very similar to the coefficient function of $\gamma^* g$ fusion. The part of the splitting function of (inconsistent) infrared origin gives a contribution, $\Delta \gamma$, to the anomalous dimensions, $\gamma$, --- it therefore exponentiates during the evolution, which is especially long in the low $x$ region. In terms of the anomalous dimensions, the inconsistency $\Delta \gamma$ is strongly enhanced in the low $x$ region by the small value of the analytic continuation of the Mellin moment $\omega$, recall $\Delta\gamma \sim {\rm const.}/\omega$. Compensation is clearly not possible over a long interval of $Q^2$.

Another good, related, example is evolution in QED (where there is 
no confinement) and where the electron density is well defined.
Here the evolution was first considered by Gribov and Lipatov \cite{GL} in terms of the Leading Log Approximation 
(rather than the `factorization theorem' approach), which allows one to trace each step of the evolution and the contribution of each Feynman diagram explicitly.
 As there is no confinement, the
electron and photon fields are well defined. Thus the only possibility
to re-define the `partons' is a field renormalization $Z(q^2)$ 
factor which does not depend on the momentum fraction $z$.
   Of course, one may consider some mixture of electron and photon PDFs,
but we discuss the original electron and photons.
  For these distributions the only way to avoid double counting 
is the `physical' approach. 
 We emphasize that it is useful to first consider the 
problem of double counting in QED where everything is well defined.
After this it is straightforward to discuss what happens in QCD.

A more general observation is that in our approach we work with partons defined by OPE operators, and the corresponding Feynman diagrams. For these quarks 
and gluons we evaluate the Feynman diagrams in both the `conventional' and the `physical' approach. 
In either case, we first have to calculate the splitting and the coefficient functions for the normal (OPE) quarks and gluons (for which we have Feynman rules and where each line in the diagrams has definite quantum numbers). Only after this may we define the new $q'$ and $g'$, and new splitting (and coefficient)  functions corresponding to an alternative factorization scheme. So differences in the infrared treatments cannot be attributed to a scheme change.

We conclude that differences in parton behaviour that follow from an analysis using a `physical' treatment of the infrared domain cannot be reproduced by a factorization scheme transition from the conventional PDF analyses. (The differences are expected to be mainly in the gluon at low scales.) The only way to account for the appropriate treatment of the infrared region is to perform a new global analysis using a complete set of corrected coefficient and splitting functions.

\section*{Appendix: Infrared contributions to NLO splitting functions}
In the axial gauge the only infrared singularity of a splitting function is due to the ladder-type (box) diagram (plus a self-energy contribution $\propto \delta(1-z)\int_0^1 dzP(z)$ which can be determined from this diagram in usual way based on the flavour and energy-momentum conservation laws).
The singularity has a logarithmic form $dk^2/k^2$ and is exactly equal to the contribution $P^{\rm LO}\otimes P^{\rm LO}$ generated by LO DGLAP evolution, since there is no other dimensionful parameter
apart from $Q^2$. To avoid double counting, we have to subtract 
 the LO DGLAP-generated contribution from the NLO splitting 
 $P^{\rm NLO}$.  This subtraction eliminates the infrared singularity 
of $P^{\rm NLO}$. However the remaining part may still contain some 
non-singular contribution of infrared (large distance) origin.

Recall that  DGLAP evolution is performed starting from some relatively large virtuality $Q^2_0>\Lambda^2_{\rm QCD}$. Everything below $Q_0$ is collected in the phenomenological `input' parton distributions, PDF($Q^2_0$). It would be best to calculate the NLO box diagram also
starting the integral over $k^2$ from $Q^2_0$; that is to consider $\int_{Q^2_0}dk^2/k^2$. On the other hand, in such a case we will introduce a new dimensionful parameter which will strongly complicate the logarithmic DGLAP evolution. If we assume that the starting scale is much smaller than the final factorization scale $\mu_F^2=Q^2$, that is $Q_0^2 \ll Q^2$, then we may consider the {\em non-singular} contribution from the region with $k^2<Q^2_0$ as a power, ${\cal O}(Q^2_0/Q^2)$, correction, and neglect it, together with the other power corrections. That is to consider the $\int_0 dk^2/k^2$.

The problem is that within the conventional approach we deal with {\em singular} integrands. Using the dimensional, $\epsilon$, 
regularization, this results in a $1/\epsilon$ pole plus some
constant terms (with respect to $\epsilon$) of infrared, $\epsilon/\epsilon$, origin. Finally we subtract the `anomalous dimension' function\footnote{See eq.(60) of \cite{Alt2}.} generated by the LO DGLAP evolution; that is we eliminate the $1/\epsilon$ pole. However, this prescription does not cancel the constant terms (with respect to $\epsilon$) of infrared, $\epsilon/\epsilon$, origin.

If, on the other hand, we adopt the `physical' approach and integrate the non-singular expression given by the box diagram minus the DGLAP-generated $P^{\rm LO}\otimes P^{\rm LO}$ contribution (or put some infrared cutoff $q_0$ on the integral
$\int_{q^2_0}dk^2/k^2$) we will never obtain such terms of infrared origin, even for $q_0\to 0$.
We thus conclude that the above LO DGLAP-generated subtraction will give the correct NLO splitting functions, and that the term coming from very large distances (greater than ${\cal O}(1/\Lambda_{\rm QCD}$)) in the `conventional' procedure should not be there.

\section*{Acknowledgements}
We thank Stefano Catani, James Stirling, Robert Thorne and Andreas Vogt for discussions.
EGdO and MGR thank the IPPP at the University of Durham for hospitality. This work was supported by the grant RFBR 11-02-00120-a
and by the Federal Program of the Russian State RSGSS-65751.2010.2;
and by FAPESP (Brazil) under contract 2011/50597-8.

\end{document}